\documentclass[pra,twocolumn,showpacs,preprintnumbers,amsmath,amssymb]{revtex4}
\usepackage{amssymb,amsmath}
\usepackage{color}
\usepackage{graphicx}% Include figure files
\usepackage{dcolumn}% Align table columns on decimal point
\usepackage{bm}% bold math
\usepackage{latexsym,epsfig}

\begin{document}
\title{ Mott  metal-insulator transition in the Doped Hubbard-Holstein model}  
%\title{Density-Driven Mott Transition in the Doped Hubbard-Holstein model}

\author{Jamshid Moradi Kurdestany  and S. Satpathy}
\email{jmkurdestany@gmail.com}
 \affiliation{Department of Physics \& Astronomy,
University of Missouri, Columbia, MO 65211, USA}

\date{\today}
\begin{abstract}
Motivated by the current interest in the understanding of the Mott insulators away from half 
filling, observed in many perovskite oxides, we study the Mott metal-insulator transition (MIT)
in the doped Hubbard-Holstein model using the Hatree-Fock mean field theory. The Hubbard-Holstein model
is the simplest model containing both the
Coulomb and the electron-lattice interactions, which are important ingredients in the physics of
the perovskite oxides. In contrast to the half-filled Hubbard model, which always results in a single phase (either metallic or insulating),
our results show that away from half-filling, a mixed phase of metallic and insulating regions occur.
As the dopant concentration is increased, the metallic part progressively grows in volume, until
it exceeds the percolation threshold, leading to percolative conduction.
This happens above a critical dopant concentration $\delta_c$, which,
depending
on the strength of the electron-lattice interaction, can be a significant fraction of unity. 
This means that the material could be insulating
even for a substantial amount of doping,  in contrast 
to the expectation that doped holes would destroy the insulating behavior of the half-filled Hubbard model. 
%with the Nagaoka theorem, where a single hole destroys the insulating behavior of the half-filled Hubbard model. 
Our theory provides a framework for the understanding of the
density-driven metal-insulator transition observed in many complex oxides.
\end{abstract}
\maketitle

\section {Introduction}

It is well known that the half filled Hubbard model is a Mott insulator\cite{Mott} when the 
strength of the on-site Coulomb interaction $U$ exceeds a critical value. 
Within the Hubbard model, the Mott insulating state can exist only at half filling, and just a single hole is supposed to destroy
the antiferromagnetic insulating ground state, turning it into a ferromagnetic metal as suggested by the Nagaoka Theorem\cite{Nagaoka}, strictly true in the infinite $U$ limit.

Quite early on, the Mott insulator LaTiO$_3$ was thought to be a prototypical example of the Nagaoka Theorem, where the undoped LaTiO$_3$ is an antiferromagnetic insulator, as predicted for the half-filled Hubbard model, but both the antiferromagnetism as well as the insulating behavior are quickly destroyed with the introduction of a small number of holes via the addition of extra oxygen\cite{Taguchi} or via $Sr$ substitution (with as little as $x \approx 0.05$ for La$_{1-x}$Sr$_x$TiO$_3$)\cite{Tokura}. Indeed, a large number of perovskite oxides have since been found to turn into metals upon hole doping, but only after a substantial amount of hole concentration has been introduced into the system. At the same time, scanning tunneling microscopy images of these doped oxides show mixed phases in the nanoscale, meaning that there is no clear phase separation with a single boundary separating the two phases, but rather that the two
phases break into intermixed nanoscale puddles. In addition, transport measurements follow percolative scaling laws with doping and temperature, further confirming the existence of the mixed phase\cite{SWCheong, Schneider, Mydosh}.

% From a theoretical point of view, there have been many studies of the phase diagrams and possible phase separation of the doped Mott insulator, with the results  depending on the precise model considered and the technique used \cite{Kotliar96, XGWen, Kotliar86, Onoda, Tremblay2009, Emery}. 
 From a theoretical point of view, there have been many studies of the doped Mott insulators
 \cite{Kotliar96, XGWen, Kotliar86, Onoda, Tremblay2009, Dagotto2, Becca, Gang, Cosentini, Arrigoni, Zitzler, Aichhorn, 
 Visscher, Andriotis, Emery,Hellberg,Gimm, Luchini,Shih, Igoshev, Balents}, 
 largely for models in two dimensions (2D), because numerical methods such as Quantum Monte Carlo are more feasible there. 
 However, the results vary depending on the methods used.
 In the 2D Hubbard model, results from quantum Monte Carlo calculations\cite{Dagotto2, Becca}  found no evidence for phase separation, consistent with the ``somewhat" exact results of Su\cite{Gang}. However, other authors using the fixed-node quantum Monte Carlo method\cite{Cosentini} or the Hartree-Fock mean-field approximation\cite{Arrigoni} have suggested phase separation in large regions of the parameter space. Phase separation at small doping levels was also found in the dynamical mean field calculation\cite{Zitzler} and the variational cluster perturbation theory works \cite{Aichhorn}. There are much fewer studies of the phase separation for the Hubbard model in 3D, although the existence of the phase separation there was suggested by the early works of Visscher\cite{Visscher} in the 1970s. The recent Hartree-Fock calculations in 3D\cite{Igoshev} and the dynamical mean-field theory (DMFT) work\cite{Millis}, strictly valid for infinite dimensions, have found phase separation in a large region of parameter space, as did the work of Andriotis et al.\cite{Andriotis}, who used the coherent-potential approximation and the Bethe lattice.
 
Phase separation in the closely related $t$-$J$ model has also been investigated because of its relevance to the cuprate superconductors. 
The phase separation has been reported for all values of $J/t$ by several authors\cite{Emery,Hellberg,Gimm}, while some authors
find it only for larger values of $J/t$ \cite{Luchini,Shih}.  There is thus  a general consensus for the phase separation in the $t$-$J$ model with a large $J/t$ and the non-half-filled band, where the system  separates into two regions, viz., an undoped antiferromagnetic region and a carrier-rich ferromagnetic region. 

All these theoretical works do not include the coupling of lattice to the electrons, which is an important ingredient in the physics of many
perovskite oxides, where a strong Jahn-Teller coupling plays a critical role in the behavior of the material. In this paper, we study the
Hubbard-Holstein model with the Hartree-Fock method, which includes both the Coulomb interaction as well as the electron-lattice coupling.
We study the energetics of the various magnetic phases including the paramagnetic and the spiral phase (which incorporates the AFM and FM phases as special cases) and compute the phase stability 
 in the doped system near half filling. For small number of dopants (electrons or holes), the system phase separates into an undoped
 antiferromagnetic insulator and a carrier-rich, ferro or spiral magnetic, metallic phase. As the dopant concentration is increased, the metallic
 part grown in volume, and eventually at a critical dopant concentration, the percolation threshold is reached and the system becomes a conductor. The critical concentration for this percolative Mott metal-insulator (MIT)  transition is studied for varying interaction parameters, and  the theoretical results are connected with the 
 existing experiments in the literature.

\section{Model}

We consider the Hubbard-Holstein model for a cubic lattice 
%\begin{widetext}
\begin{eqnarray}
 {\cal H} &=&  \sum_{ \langle ij \rangle \sigma} t_{ij} (c_{i\sigma}^{\dagger} c_{j\sigma} + H.c.)
+U\sum_i { n}_{i\uparrow}{ n}_{i\downarrow} \nonumber \\
&+& \sum_{i}(\frac{1}{2}KQ^2_i - gQ_i n_i)-\mu\sum_{i \sigma}n_{i\sigma},
\label{HHmodel} 
\end{eqnarray}
%\end{widetext}
which contains both the Coulomb interaction and the electron-lattice coupling terms. 
Here  $c_{i\sigma}^{\dagger}$ is the electron creation operator at site $i$ with spin $\sigma$, $n_{i\sigma} = c^{\dagger}_{i\sigma}c_{i\sigma}$
is the number operator, $t_{ij}$ is the hopping amplitude between nearest-neighbor sites denoted by $\langle ij \rangle$, $U$ is the onsite Coulomb repulsion,  $Q_i$ is the lattice distortion at site $i$, $K$ and $g$ are, respectively, the stiffness  
and the electron-lattice coupling constants, and $\mu$ is the chemical potential that controls the carrier concentration. 
Taking the nearest-neighbor hopping integral as $t_{ij} = -t$, there are two parameters in the Hamiltonian, viz., $U/t$ and 
$\lambda \equiv g^2/ ( KW)$, where $ W = 12 |t|$ is the band width and $\lambda$  is the effective electron-lattice coupling strength. Note that we have considered the static Holstein model\cite{Holstein}, which  contains a simpler version of the local lattice interaction such as the Jahn-Teller interaction, and, in addition, it does not contain any phonon momentum dependence.

The key problem to study is the energy of the ground state and the stability of the various phases as a function of the 
carrier concentration away from the half filling.
Both magnetic (ferro, antiferro, or spiral) as well as non-magnetic phases are considered.
In fact, all these solutions are special cases of the spiral phase, which is conveniently described in terms of a 
site-dependent local spin basis set described by the unitary transformation\cite{Arrigoni} 
\begin{equation}
d^{\dagger}_{i\sigma} = 
\sum_{\sigma ^\prime}  (e^{-i\vec{\sigma} \cdot \vec{\alpha_i}/2})_{\sigma \sigma ^\prime} \   c^{\dagger}_{i\sigma ^\prime},
\label{d-dagger}
\end{equation}
where  $\vec{\alpha_i}$ is the site dependent spin rotation angle. 
The spiral phase is described by
$\vec{\alpha_{i}}=(\vec{q}  \cdot  \vec{R_i}) ~ \hat{x}$, where $\hat{x}$ is the spin rotation axis, $\vec R_i$ is the site position, and $\vec q \equiv (q_x, q_y, q_z)$  is the modulation wave vector of the spiral state. The ferro, para, as well as the antiferromagnetic states, considered in this work, are all special cases
of the spiral state. Explicitly,  $\vec q = 0$ for the ferro or paramagnetic state, while it is
$\pi (1, 1, 1)$ for the N\'eel    antiferromagnetic state.

In the new basis, the Hamiltonian (\ref{HHmodel}) remains unchanged except for the first term, which becomes
${\cal H}_{ke}  =  \sum_{\langle ij \rangle, \sigma\sigma^\prime  }      (t^{\sigma\sigma^\prime}_{i j}d^{\dagger}_{i\sigma}d_{j\sigma^\prime}+H.c.) $, 
where the hopping is now spin-dependent
\begin{equation}
t^{\sigma\sigma  ^ \prime}_{i j}  = (e^  {i\vec{q} \cdot (\vec{R_i}-\vec{R_j})\sigma_{x}/2})_{\sigma\sigma^ \prime} t_{i j}, 
\end{equation}
and in the remaining terms in (\ref{HHmodel}), the number operators are redefined to mean $n_{i\sigma} = d^{\dagger}_{i\sigma}d_{i\sigma}$.
Making the Bloch transformation into momentum space
\begin{equation}
d^{\dagger}_{\vec k\sigma} =  \frac{1}{\sqrt{N}} \sum_{i}e^{i\vec k.\vec{R_i}}d^{\dagger}_{i\sigma},
\end{equation}
and using the Hartree-Fock approximation:
$  n_1 n_2 = \langle n_1\rangle n_2+   \langle n_2\rangle n_1  
 - \langle d_1^\dagger d_2\rangle  d_2^\dagger d_1
  -\langle d_2^\dagger d_1\rangle  d_1^\dagger d_2 
 -  \langle n_1\rangle  \langle  n_2\rangle   + \langle d_1^\dagger d_2\rangle  \langle d_2^\dagger d_1\rangle$, 
%
%\begin{eqnarray}
%  n_{1} n_{2} &=& \langle n_{1}\rangle n_{2}+\langle n_{2}\rangle n_{1}  - \langle c_1^\dagger c_2\rangle  c_2^\dagger c_1
%  -\langle c_2^\dagger c_1\rangle  c_1^\dagger c_2 \nonumber   \\
%  &-& \langle n_1\rangle  \langle  n_2\rangle   + \langle c_1^\dagger c_2\rangle  \langle c_2^\dagger c_1\rangle ,  
% \end{eqnarray}
we get the quasi-particle Hamiltonian
\begin{equation}
{ \cal H} (\vec k) = 
\left[ 
{\begin{array}{*{20}c}
   T_{1}(\vec k)+U\langle  n_{\downarrow} \rangle-\mu &    \     -T_{2}(\vec k) - U \langle d_\downarrow ^\dagger d_\uparrow\rangle    \\
  - T_{2}(\vec k)  - U \langle d_\uparrow ^\dagger d_\downarrow \rangle &          \ \ \                T_{1}(\vec k)+U \langle n_{\uparrow} \rangle- \mu  \\
 \end{array} } 
 \right],
 \label{Hk}
\end{equation}
where 
$ T_1  (\vec k) = -   2t  [ \cos(k_x a) \cos(q_x a / 2) + \cos(k_y a)    \cos(q_y a / 2) + \cos(k_z a) \cos(q_z a / 2)]  $, 
$ T_2  (\vec k) $ is the same as $ T_1  (\vec k) $ except that all cosine functions are replaced by sines, only the nearest-neighbor hopping $t_{ij} = -t$ has been kept in the original Hamiltonian (the unit of energy is set by $t = 1$), and the expectation values
$\langle d^\dagger_\sigma d_{\sigma ^\prime} \rangle $ are to be determined self-consistently.
Note that the exact form of ${\cal H}_k$ would depend on the spin rotation axis $\vec \alpha$ in the spiral phase (here chosen along $\hat x$).
However, the final results should not depend on this choice as there is no coupling between the space and the spin coordinates. 
We also find from direct calculations that the exchange terms 
$U \langle d_\sigma ^\dagger d_{-\sigma}\rangle$
 appearing  in Eq. (\ref{Hk}) contribute very little to the total energy. 
 This contribution would be exactly zero, if the spins don't mix, so that the density matrices
 $\rho_{\sigma \sigma^\prime} \equiv  \langle d_\sigma ^\dagger d_{\sigma^\prime} \rangle$
  are diagonal in the spin space.
 
 The total energy per site is given by
\begin{equation}
E(\vec q) = \frac{1}{N} \sum^\mu _{\vec k \sigma }  \varepsilon_{\vec k \sigma} 
- U\langle n_{\uparrow}\rangle \langle n_{\downarrow}\rangle 
+  U \langle d_\uparrow^\dagger d_\downarrow \rangle   \langle d_\downarrow^\dagger d_\uparrow \rangle - \frac {g^2 n^2} {2K} ,
\label{Eq}
\end{equation} 
where $ \varepsilon_{\vec k \sigma} $ are the eigenvalues of  the Hamiltonian in Eq. (\ref{Hk}),
the second and the third terms correct for the double counting of the Coulomb energy, and the last term is the lattice energy gain at each site, obtained from minimizing the lattice energy $ \partial E /  \partial Q = 0$ from Eq. (\ref{HHmodel}). The chemical potential is related to the number of electrons by the expression
$ N^{-1} \sum_{\vec k \sigma } \theta (\mu - \varepsilon_{\vec k \sigma}) = n$, $N$ being the number of lattice sites. For a fixed value of doping $\delta = 1-n$, where $n$ is the total number of electrons per lattice site, we have minimized the total energy $E (\vec q)$ numerically as a function of the spiral vector $\vec q$ by varying each component between $0$ and $2 \pi$. The minimum yields the ground state. All Brillouin zone integrations were performed with 1000 $k$-points. We restrict ourselves to the hole doping region $n \le 1$ without loss of generality, since we have the electron-hole symmetry in the problem.
%

%: Fig 1
\begin{figure}[tbp]
\centering \includegraphics[width=7.0cm]{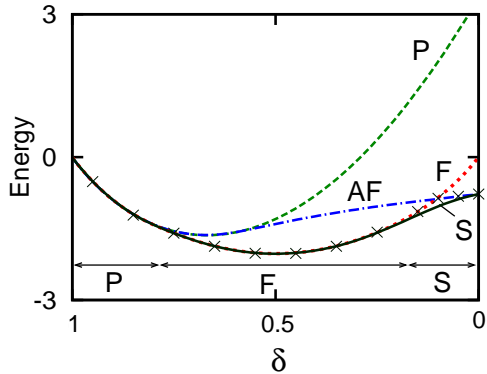}
\caption{Ground-state energy as a function of the hole concentration $\delta$ for $U/W = 1.25$ and $\lambda = 0$. The lines show results for a single-site unit cell, while the crosses show the results for a double-site unit cell,
which allowed for  charge and spin disproportionation, but no such disproportionation was found, and the double-site results converged to the single-site results. Note that the double-site calculations allowed for the ferro or the anti-ferro phase, but not the spiral phase.
}
\label{fig1}
\end{figure}

\section{Results }

To determine the phase diagram, we calculated the ground-state energy of the system according to Eq.(\ref{Eq}) for the given input parameters $n$, $U$, and $\lambda$. Figure (\ref{fig1}) shows a typical plot of the ground-state energy per lattice site as a function of the hole concentration $\delta$ for different magnetic phases. As seen from the figure, the ground state is antiferromagnetic (AF) at half-filling ($\delta=0$), in agreement with the standard result for the Hubbard model. With increasing hole concentration $\delta$, the system first turns into a spiral (S) state, then into a ferromagnetic (F) state, and  eventually into the paramagnetic (P) state. 

Note that we have considered the spiral state in Eq. (\ref{d-dagger}), which is a spin density wave (SDW) state, with the modulation wave vector $\vec q$, but not the charge density wave (CDW) state, which is a higher energy state and is not expected to occur in the parameter regime we are working. The CDW state is difficult to incorporate within our calculation as it
requires a supercell of arbitrary size depending on the modulation wave vector of the CDW. However, we can study the CDW in a special case, viz., where  the modulation $\vec q = (\pi, \pi, \pi)$, in which case we have two sites in the unit cell of the crystal, and we can allow for both charge and spin disproportionation between the two sublattices. Results of this calculation are also shown in Fig. (\ref{fig1}) as crosses and they go over to the single-site results indicating the absence of any CDW for this wave vector.

We note further that the CDW state could be favored when the electron-lattice interaction is strong. We can estimate the condition for this by considering the energy of the charge-disproportionated state (a special case of the CDW) for the half-filled
Hubbard-Holstein model and comparing it with the energy of the state without any charge disproportionation. In the former case, the charges on the two sublattices are $1 \pm \eta$ ($\eta \le 1$ is the charge disproportionation amplitude), and the total energy would be $E = -(g^2/2K) [(1+\eta)^2 + (1-\eta^2)] + U \eta$. 
The first term here is the energy gain due to lattice interaction, the second term is due to the fact that $\eta$ electrons are forced to occupy the upper Hubbard band, and we have neglected the kinetic energy difference
in order to get a simple estimate.
 It immediately follows from this expression that such a CDW state 
would be favorable if $U/(W \lambda) \le 1$. For the parameter regime relevant for the oxides, this condition is not satisfied, so that it is reasonable to omit the CDW state, which we have not considered in our work.
%Note that if we ignore the CDW,  the phase diagram, Fig. ({\ref{fig2}), does not depend on the electron-lattice coupling strength $\lambda$, as it affects the energy of all phases equally for a fixed number of electrons $n$.

%: Fig 2
\begin{figure}[tbp]
\centering \includegraphics[width=7.0cm]{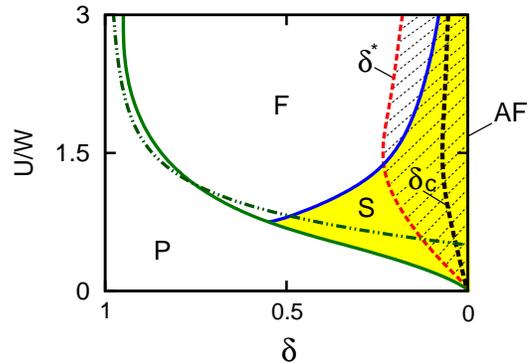}
%\centering \includegraphics[width=7.0cm]{fig2-old.eps}
\caption{Ground-state phase diagram for the Hubbard model for the simple cubic lattice. The red dashed line separates the stable and unstable single phase regions, while the black dashed  line indicates the MIT (these two lines were calculated from the total energy curve $E(\delta)$ for each $U/W$ as illustrated in Fig. (\ref{fig3}) below).  The hatched region indicates existence of the mixed phase. The dash dotted line shows the Stoner criterion result, Eq. (\ref{PF-line}), for a sinusoidal 
model density-of-states.
}
\label{fig2}
\end{figure}

Fig. ({\ref{fig2}) shows the calculated phase diagram.  For the half-filled case, there is perfect Fermi surface nesting 
[$\varepsilon (\vec k_F) = \varepsilon (\vec k_F + \vec q_n$) = 0, $\vec q_n = \pi (1, 1, 1)$ is the nesting vector], 
which leads to an anti-ferromagnetic insulator for any value of  $U$. 
As we move away from half filling, perfect nesting is lost and a critical value $U_c$ is needed for the onset of magnetic order. Below a certain hole doping $\delta$, the system goes from the paramagnetic state to a spiral state, and eventually to the ferromagnetic state,
as $U$ is increased, while for a larger value of $\delta$, the system goes directly from the paramagnetic to the ferromagnetic state.   

Fig. (\ref{Fig-E-vs-q}) shows the calculated energy as a function of the spiral wave vector for three different parameters. Note that for the paramagnetic solution corresponding to $U/W = 0.3$, the energy is independent of the spiral wave vector, since the magnetic moment is zero.

%: Fig3
\begin{figure}[tbp]
\centering \includegraphics[width=7.0cm]{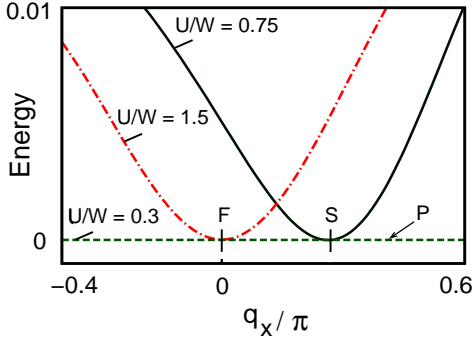}
\caption{Energy as a function of the spiral wave vector $\vec q = (q_x, \pi, \pi)$  for the Hubbard
model with three different Coulomb parameters, indicating the paramagnetic (P), ferromagnetic (F), or the 
spiral (S) ground state, depending on the strength of the Coulomb U.  Here, the hole concentration is $\delta = 0.6$ and the zero of the energy has been shifted to correspond to the minimum in each case.
}
\label{Fig-E-vs-q}
\end{figure}

%:Para-Ferro Boundary
{\it Para-Ferro phase boundary} -- The boundary between the paramagnetic and the ferromagnetic phases in the Hubbard model (Fig. \ref{fig2}) can be understood by taking a model
density of states and applying the Stoner criterion for ferromagnetic instability. We consider the sinusoidal density-of-states for each spin
\begin{eqnarray}
\rho (\varepsilon) =\begin{cases}
\frac{\pi}{2 W} \sin \ (\varepsilon \pi / W) \hspace{3mm}  \text { if  $ 0  < \varepsilon < W $},   \\
0  \hspace{25mm}  \text { else},
\end{cases}
\label{model energy}
\end{eqnarray}
of bandwidth $W$. The total energy $E$ is a sum of the band energy, the Coulomb energy, and the lattice energy, which is immediately obtained from a direct integration to yield
\begin{eqnarray}
E(n,m) &=& \frac{W}{2 \pi}  \big[ \sqrt{1-x^2} - x \cos^{-1} x   \nonumber \\
+ \sqrt{1-y^2} &-&  y \cos^{-1} y \big ] +\frac{U}{4} (n^2 - m ^2) -\lambda n^2,
\label{E-nm}
\end{eqnarray}
where  $x = 1-n-m$, $y=1-n+m$, $n =n_\uparrow + n_\downarrow$ is the number of electrons, and $m =n_\uparrow - n_\downarrow$ is the
spin polarization. The Fermi energies for the up and down spins are, respectively, $\varepsilon _{F\uparrow} = \pi^{-1} W \cos^{-1} x$ and 
$\varepsilon _{F\downarrow} = \pi^{-1} W \cos^{-1} y$. The onset of  ferromegetism is determined from the Stoner criterion $U \rho (\varepsilon_F) \ge 1$, where $ \varepsilon_F = \pi^{-1} W \cos^{-1} (1- n)$ is the Fermi energy of the paramagnetic phase, while the spin polarization is determined by the minimization of the energy, Eq. (\ref{model energy}), as a function of the polarization $m$. The  Stoner criterion leads to the equation of the para-ferro transition line: 
\begin{equation}
\delta  = 1-n = \sqrt {1- (\frac{2W}{\pi U})^2},
\label{PF-line}
\end{equation}
which is plotted as a dotted line in Fig. (\ref{fig2}) and reproduces the trend found from the full solution of the Hubbard model for the cubic lattice. It is readily seen from Eq. (\ref{PF-line}) that for the Coulomb interaction below the critical value $U_c = 2/ \pi$, the system is paramagnetic for all values of the hole concentration $\delta$.
%Note that if we ignore the CDW, the phase line (and in fact the entire phase diagram) does not depend on the electron-lattice coupling strength $\lambda$, as it affects the energy of all phases equally for a fixed number of electrons $n$.

%: Fig 4
\begin{figure}[tbp]
\centering \includegraphics[width=7.0cm]{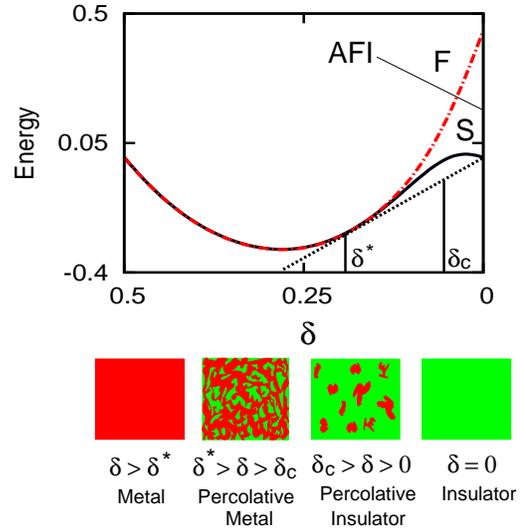}
\caption{Ground-state energy  as a function of the hole concentration $\delta$, indicating phase instability near half filling ($\delta = 0$), and
the Maxwell construction that yields
the upper concentration $\delta^*$ for the existence of the mixed phase. The system remains an insulator for $\delta < \delta_c$, the percolation threshold, beyond which the metallic fraction forms a percolation network making the system a conductor.
Parameters used are: $U/W = 2.0 $   and $\lambda = 0$.}
\label{fig3}
\end{figure}

%: Fig 5
\begin{figure}[btp]
\centering \includegraphics[width=7.0cm]{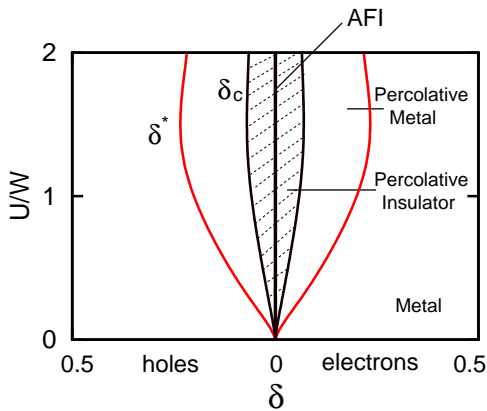}
\caption{Phase diagram indicating the various phases as a function of carrier (electron or hole) doping $\delta$. 
As $\delta$ is increased starting from the AF insulator state at half-filling ($\delta = 0$), the system continues to be a mixed-phase insulator, turning into a percolative metal beyond $\delta = \delta_c$, and eventually becoming a single-phase metal beyond $\delta^*$ as discussed in the text. 
}
\label{fig4}
\end{figure}

{\it Percolative metal-insulator transition} --
Returning to our original Hubbard-Holstein model, as seen from Fig. (\ref{fig1}), the ground-state energy is {\it not}  everywhere convex, which  indicates a phase separation, which is seen for small doping near half filling. At half filling, we have an antiferromagnetic insulator. As holes are introduced, the system phase separates
into two regions, one is the anti-ferro insulating state with hole concentration zero, and the second is a spiral or ferro phase (depending on the strength of $U$) with hole concentration $\delta^*$. As $\delta$ is increased, so does the volume of the metallic fraction. When it exceeds a certain threshold $\delta_c$, given by the percolation theory, the metallic regions form a percolative network and the system conducts.

The fraction of the two phases can be obtained from the standard Maxwell construction,
which is illustrated for the case of $U/ W = 2$ in Fig. (\ref{fig3}).
If $v_m (v_i)$ is the volume fraction of the substance in the metallic (insulating) phase in the mixed phase region ($\delta^* < \delta < 0$), 
then we have the two equations: $v_m  + v_i = 1$ and $v_m \delta^* = \delta$, which means that the metallic volume fraction linearly
increases with the hole concentration, i.e., $v_m  = \delta  /  \delta^*$. 
The hole concentration $\delta^*$ separates the mixed phase region from the single phase region and depends on the Hamiltonian parameters as seen, e.g., from Fig. (\ref{fig2}), and it must be calculated from the total energy curve for each set of parameters from a Maxwell construction.

The Maxwell construction indicates phase separation into two separate regions consisting of single phases, separated by a single boundary. However, in the actual solids, one does not encounter such clear phase separation, but rather a mixed phase usually results,
where the two phases are intermixed on the nanoscale. 
There are many reasons why a mixed phase could be more favorable.
For example, the presence of a small amount 
of charged impurities because of unintentional doping could cause a deviation from charge neutrality of the two components and would impede the formation of the phase separation due to the large cost in Coulomb energy. 
Thus one would encounter a nanoscale inhomogeneous phase (or mixed phase) with intermixed metallic and insulating components (Coulomb frustrated phase separation)\cite{HRK}. It has also been suggested that the mixed phase could even originate due to kinetic reasons, i.e., self-organized inhomogeneities resulting from a strong coupling between electronic and elastic degrees of freedom \cite{KHAHN}. A large number of experiments point to the existence of the mixed phases in the oxide materials, including transport results and scanning tunneling microscopy images.\cite{Loa,Baldini,Ramos}

The percolation threshold $v_c$, beyond which the metallic regions touch and the percolative conduction begins, depends on the
specific model used in the percolation theory, but is typically about $v_c \approx 0.30$. 
For example, in the percolation model, where the metallic region consists of randomly-packed, overlapping spheres
of radius $r$ in an insulating matrix,  the critical volume fraction of the spheres for the onset of percolation  is  $v_c \approx 0.29$ and is independent of $r$ \cite{Pike}.  
On the other hand, for the site percolation problem in the cubic lattice, the percolation threshold is about $v_c \approx 0.31$.  The site percolation thresholds are long well known,\cite{Efros} but are summarized in Table I for ready reference. We have used the value $v_c = 0.3$ in our calculations, which is similar to the site percolation
result for the cubic lattice.

%==============================================
%:Table 1:  Percolation Thereshold

\begin{table}[h!]
\centering
\begin{tabular}{|c| c| c| c| c| c| c| c|} 
\hline
 & cubic & diamond & bcc & fcc & square & triangular& honeycomb \\ 
\hline
$v_c$ & 0.31 & 0.43 & 0.25 & 0.20 & 0.59 & 0.5 & 0.7 \\ 
\hline
\end{tabular}
\caption{Site percolation threshold $v_c$ for various lattices}
\label{table1}
\end{table}
%==============================================

Percolative conduction occurs, when the metallic volume fraction exceeds $v_c$, 
i.e., $\delta/ \delta^* = v_m > v_c$, or
\begin{equation}
\delta  >  \delta_c = v_c \delta^*, 
\label{delta-c}
\end{equation}
where $\delta^*$ is the critical concentration, beyond which the system turns into a single-phase metal, which is either ferromagnetic
or in the spin spiral state depending on the strength of $U/W$ (see Fig. (\ref{fig2})). For the specific parameters used in Fig. (\ref{fig3}), the full metallic phase for $\delta > \delta^*$ is ferromagnetic; 
for intermediate values of $\delta$ between $0$ and $\delta^*$, the phase separation occurs between the AFI half-filled 
($\delta = 0$) phase and the FM metallic phase with carrier concentration $\delta^*$.
Fig. (\ref{fig4}) summarizes the  phase diagram showing the MIT boundary. The system continues to remain an AF insulator until dopant concentration (electrons or holes) exceeds the critical value $\delta_c$. 

 %
%==============================================
%On the other hand, for the site percolation problem in the cubic lattice, the percolation threshold is about $v_c \approx 0.31$.  The site percolation thresholds are long well known,\cite{Efros} but are summarized in Table I for ready reference.

%:Table 1-- Do not include

%\begin{table}[h!]
%\centering
%\begin{tabular}{|c| c| c| c| c| c| c| c|} 
%\hline
% & cubic & diamond & bcc & fcc & square & triangular& honeycomb \\ 
%\hline
%$v_c$ & 0.31 & 0.43 & 0.25 & 0.20 & 0.59 & 0.5 & 0.7 \\ 
%\hline
%\end{tabular}
%\caption{Site percolation threshold $v_c$ for various lattices}
%\label{table1}
%\end{table}
%==============================================
%

%:Effect of electron-lattice coupling
{\it Effect of electron-lattice coupling} -- A finite value of the electron-lattice coupling in the Hubbard-Holstein model does not change the relative energies of the various phases for a fixed
concentration $n$ as already noted, since it alters the energy of each phase equally (see Eqs. (\ref{Eq}) and (\ref{E-nm})). 
The presence of charge disproportionation or a CDW ($n$ varies from site to site) would change the phase diagram; However, as we have already argued at the beginning of this Section, for parameters relevant to the oxides, the CDW phase is unlikely to occur,
which we have not considered in this work.
Thus the various phase regions (AF, F, P, or S) in the phase diagram, Fig. (\ref{fig2}), remain unchanged. However, the curvatures of the ground-state total energy as a function of $n$ or $\delta$, as in Fig. (\ref{fig1}), change, leading to the phase separation regions which now change with $\lambda$, and therefore so do  the quantities
$\delta_c$ and $\delta^*$. This is clearly seen from Fig. \ref{fig5}, where $\delta^*$ increases as the electron-lattice coupling strength $\lambda$ is increased.

Fig. (\ref{fig6}) shows the critical doping $\delta_{c}$ as a function of the electron-lattice coupling strength $\lambda$ for several values of $U/W$.  As seen from the figure, the larger the value of $\lambda$, the higher is the dopant concentration $\delta$ needed for the transition into the metallic state. Finally, Fig. (\ref{fig7}) shows the phase diagram in the Hubbard-Holstein model for a specific value of $\lambda$.

%
%: Fig 6
\begin{figure}[bt]
\centering \includegraphics[width=7.0cm]  {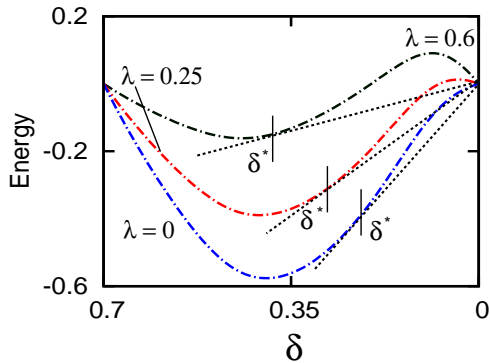} 
\caption{Energy  vs. doping $\delta$ for different strengths of the electron-lattice coupling $\lambda = 0, 0.25$, and 0.6 with $U/W=1.25$.
A linear term const. $\times n$ has been subtracted from the energy and the zero of the energy has been redefined to more clearly show the
Maxwell construction.}
\label{fig5}
\end{figure} 
%: Fig 7
\begin{figure}[th]
\centering \includegraphics[width=7.0cm]   {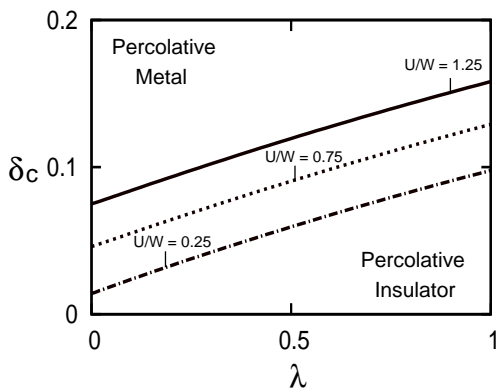}
\caption{The critical dopant concentration $\delta_c$ for the MIT  
as a function of the electron-lattice coupling strength $\lambda$. 
$\delta_c$ was obtained from the Maxwell construction (Fig. (\ref{fig5})) and Eq. (\ref{delta-c}).
}
\label{fig6}
\end{figure}
%

%: Fig 8
\begin{figure}[tb]
\includegraphics[width=7.0cm]{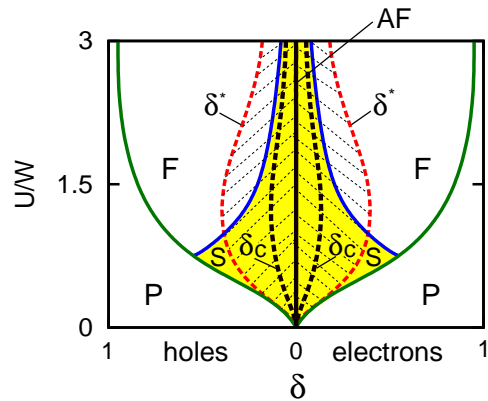}
\caption{ Ground state phase diagram for the Hubbard-Holstein model for the simple cubic lattice  for $\lambda$ = 0.6, with $\delta_c$ indicating the critical carrier concentration for percolative MIT, 
where $\delta > \delta_c$ is the metallic region. }
\label{fig7}
\end{figure} 

%
%: Table 2
%=============================== 
\setcitestyle{square}
\begin{table}[tb]
\centering
\begin{tabular}{|c| c| c| c|} 
\hline
 perovskite oxide & critical hole  & Ref. \\ 

 & doping ($\delta_c $)& \\ 
\hline
\hline
SmNiO$_{3}$ & 0.1 & \cite{SmCaNiO3} \\
\hline
LaTiO$_{3}$ & 0.05&  \cite{Tokura} \\
\hline
PrTiO$_3$ & 0.14  &  \cite{RCaTiO3} \\
\hline
NdTiO$_3$ & 0.2&  \cite{RCaTiO3} \\ 
\hline
SmTiO$_3$ & 0.24& \cite{RCaTiO3} \\
\hline
YTiO$_{3}$ & 0.35 & \cite{YCaTiO3} \\
\hline
LaMnO$_{3}$ & 0.17 - 0.2&  \cite{LaCaMnO3, LaSrMnO3} \\
\hline
PrMnO$_{3}$ & 0.3 - 0.5&  \cite{1PrCaMnO3,2PrCaMnO3, NdSrMnO3} \\
\hline
NdMnO$_{3}$ & 0.5 & \cite{NdSrMnO3} \\
\hline
LaVO$_{3}$ & 0.176& \cite{LaSrVO32006} \\
\hline
YVO$_{3}$ & 0.5&  \cite{YcaVO31993, YCdVO3} \\
\hline
\end{tabular}
\caption{ Summary of the experimental results for the critical hole doping in the perovskite oxides for transition to the metallic state. }
\label{table2}
\end{table} 
%=============================== 

To make connections with the experiments, we summarize the measured critical carrier density for the MIT in several perovskite oxides from the existing literature in Table II. As these results indicate, the critical carrier concentration $\delta_c$ needed to transform the insulating phase into the metallic phase is a significant fraction of unity, starting from 0.05 for LaTiO$_3$ to as high as 0.5 for YVO$_3$. However, other than a few systems, where $\delta_c$ is as high as 0.5, for most compounds shown in Table II, it is between 0.05 and 0.2, which is the typical value of $\delta_c$ predicted by our theory. 

Fig. (\ref{fig-expt}) shows the experimental conductivity behavior\cite{RCaTiO3} of the doped titanates RTiO$_3$ plotted against the bandwidth of the material as well as the same calculated from our theory. 
Although inclusion of the detail interactions in the Hamiltonian may be necessary for a quantitative description of a specific compound, 
the general trend for the onset of the MIT is well described within the Hubbard-Holstein model. 
As seen from Fig. (\ref{fig-expt}), for a large bandwidth ($U/W$ less than a critical value), the system is a metal for all doping levels, and as $U/W$ is increased beyond a critical value, the critical carrier concentration for MIT increases, roughly linearly. This agrees with the experimental data, where Katsufuji et al.\cite{RCaTiO3} have plotted the inverse bandwidth vs. the conduction behavior for a large number of samples with different carrier concentrations in the titanates.    
As was argued in Ref. \cite{RCaTiO3}, the magnitude of the Coulomb $U$ may be expected to be relatively unchanged 
for the R$_{1-x}$Ca$_x$TiO$_{3+y/2}$ series, allowing a direct comparison of the trends seen in theory vs. experiments.

One point to note is that Eq. (\ref{delta-c}) puts an upper limit on the  critical doping $\delta_c \approx 0.3 $, since 
$\delta^*$ can not exceed one and $v_c \approx 0.3 $, which is what is observed for most of the samples in Table I. For carrier concentration $\delta$ as high as 0.5, as is the case for some of the samples, the crystal and electronic structures are likely changed significantly, making the model less applicable for such systems. In our theory, we have assumed that the percolative conduction occurs in the mixed phase, where the two components (metallic and insulating) occur randomly, so that the percolation theory applies. If the two components do not occur randomly, but rather that there is a tendency towards coalescing of the components, this would increase the
critical value $\delta_c$, as more volume fraction of the metallic component will be needed before a percolation path for conduction forms.

Note that the Hartree-Fock approximation due to its mean-field nature does omit the  effect of fluctuations on the phase separation. It has been shown that such quantum fluctuations can indeed modify the magnetic phase boundary within the Hubbard model\cite{fluctu1,fluctu2}. 
However, the qualitative similarity of our theoretical results with the experiments (as seen from Fig. (\ref{fig-expt})) suggests that the Hartree-Fock results should contain the qualitative physics of the problem, while the fluctuation effects will likely alter the predicted critical doping quantitatively. 
The effect of the fluctuations on the phase separation remains an open question for future study.

%: Fig 9
\begin{figure} [bt]
\includegraphics[width=5.5cm]{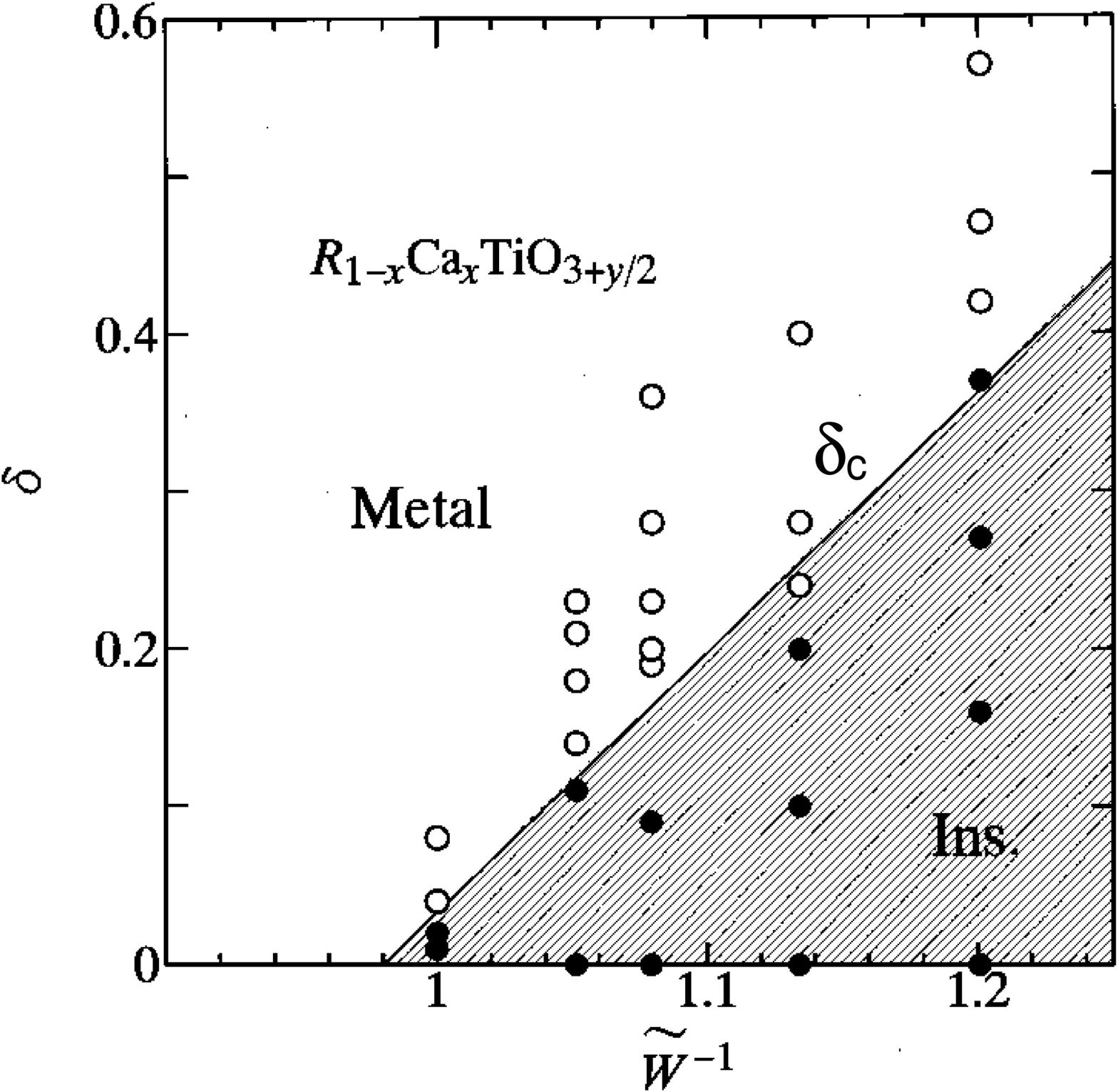} 
\includegraphics[width=6.0cm]{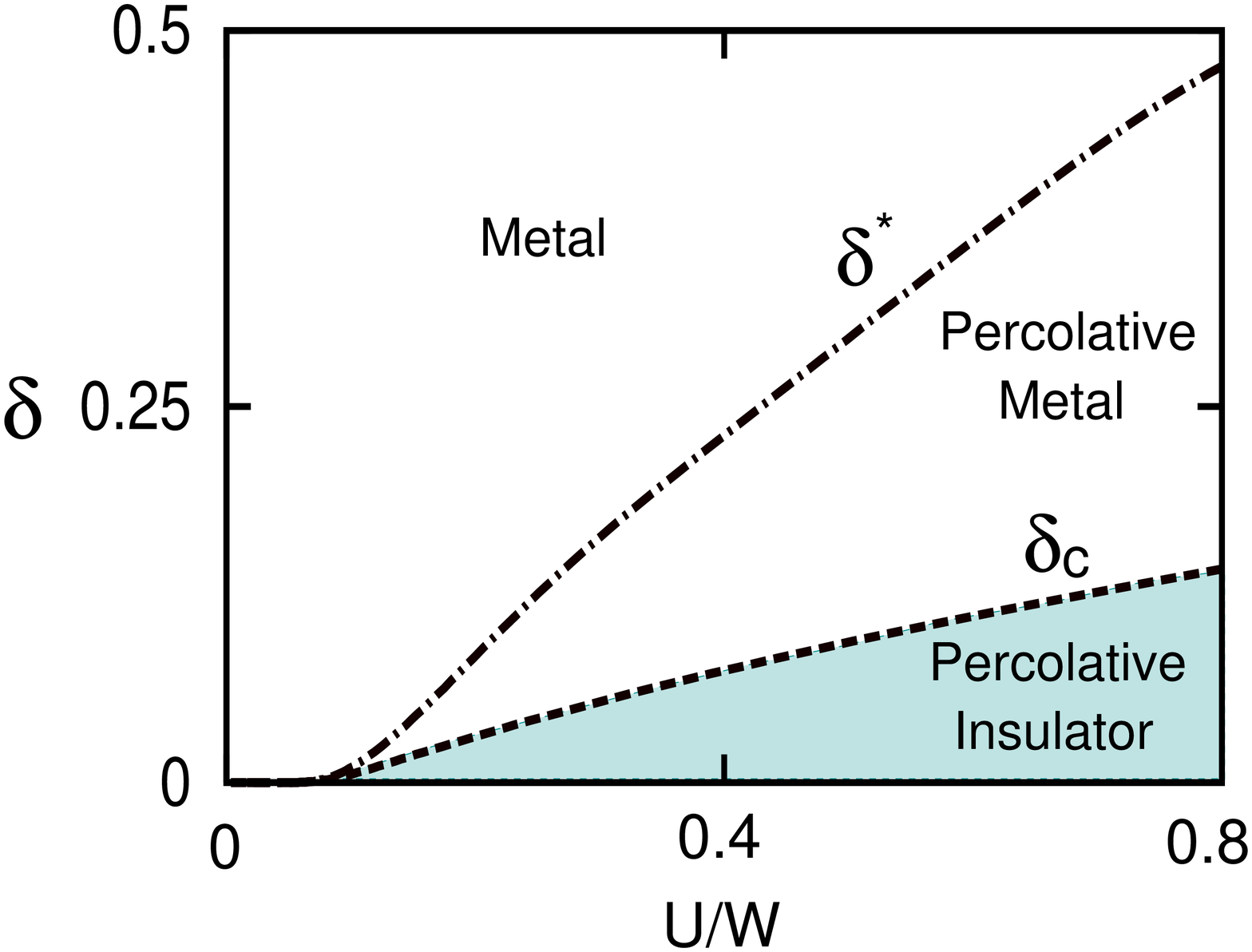}
\caption{ Experimental conductivity data, taken from
Katsufuji et al.\cite{RCaTiO3},
for the hole-doped RTiO$_3$ system, with the x-axis showing the renormalized bandwidth $ \tilde {W}$ (ratio of the bandwidth of each RTiO$_3$ to that of LaTiO$_3$)  (top) and the 
theoretical phase diagram from the present calculations, with the electron-lattice coupling strength $\lambda = 0.6$ (bottom).   
 }
\label{fig-expt}
\end{figure}

\section{Summary}

In summary, we studied the phase diagram and energetics of the Hubbard-Holstein model using the Hartree-Fock method. For a wide range of the Hamiltonian parameters, we found the existence of a mixed phase, consisting of 
an undoped component which is an anti-ferro insulator and a carrier-rich metallic phase, which is either ferromagnetic or spiral magnetic. As the carrier concentration (electrons or holes) increases with doping, the metallic portion slowly grows forming isolated islands in an insulating matrix. As the volume fraction of the metallic 
islands increases with carrier doping, eventually they form a percolative conducting network and the material conducts beyond
the critical dopant concentration $\delta_c$. This happens for $\delta_c$ which is typically between zero and 0.2 or so, in general agreement with the experimental results. 
We furthermore showed that the electron-lattice interaction favors the insulating phase with respect to the metallic phase and the critical doping value increases along with the  strength of the electron-lattice coupling. 
The general trends for the critical doping concentration for MIT predicted by our theory agrees with the existing experimental results for the hole doped perovskite oxides.

This research was supported by the U.S. Department of Energy, 
Office of Basic Energy Sciences, Division of Materials Sciences and Engineering under Grant  No. DE-FG02-00ER45818.

\end{document}